\documentstyle[preprint,eqsecnum,aps]{revtex}
\def\prb{Phys. Rev.}
\begin{document}
\draft
\preprint{CIEA-95/1 CM}
\title{
(001)-surface-induced bulk and surface states in
wide band gap zincblende II-VI semiconductors.
}
\author{D. Olgu\'{\i}n and R. Baquero}
\address{
Departamento de F\'{\i}sica, CINVESTAV, A. P. 14-740, 07000 M\'exico D.F.
}
\maketitle
\begin{abstract}
In a previous paper [\prb\ {\bf 50}, 1980 (1994)] we gave account of
the nondispersive band first found experimentally at --4.4 eV for
CdTe(001) by Niles and H\"ochst. We have characterized this band as
a surface--induced bulk state. In a second paper we showed that a similar
state does exist in II--VI and III--V zincblende semiconductor
compounds. In this paper we show that there are more such states within
the valence band energy interval.
We use tight-binding hamiltonians and the surface Green's function
matching method
to calculate the surface and surface--induced bulk states in the wide band gap
zincblende semiconductors CdTe, CdSe, ZnTe and ZnSe. We find a distinctive
surface state for the cation and two for the anion termination of the
(001)--surface
and three (001)--surface--induced bulk states with energies that correspond to
the value of the heavy hole, light hole and spin--orbit bands at $X$.
\end{abstract}

\narrowtext
\section{Introduction}
The study of the physics of surfaces, interfaces, quantum wells and
superlattices
of semiconductors has call interest in the last few
years.\cite{bryant,lowther,niles,gawlik,daniel1,bala,kilday,duc,arriaga}
The interest is
not only for binary compounds
but also, more recently, for
ternary and quaternary compounds.

The starting point is an accurate
description of the electronic band structure of the binary compound by a method
which serves as a sound basis to a clear and simple description of the
more complicated compounds and systems.

In previous work\cite{noguera,quintanar} we have used
the tight-binding formulation
to calculate the total density of states (DOS) and,
in conjunction with
the known surface Green's
function matching method (SGFM),\cite{g-m} to derive the surface
and the interface DOS. In this paper we want to use the same method to
calculate the surface and surface--induced bulk states. The method can be
also applied to superlattices\cite{arriaga,rafa-moliner}
and phonons.\cite{brito}
These band structures can be also used
to calculate transport properties in heterostructures as quantum wells, for
example, by making use of the well known many--body formulation by
Keldysh.\cite{keldysh}

In this paper, we consider the band structure of the II-VI wide band
gap semiconductors CdTe, ZnTe, CdSe and ZnSe.
We have first obtained the bulk bands (infinite medium) from the direct
diagonalization of the tight--binding Hamiltonians and from the poles of the
real part of the corresponding Green's function which is totaly equivalent.
These very same results were obtained using the SGFM method, from the
(001)--bulk--projected Green's function. This is actually a proof of
consistency which gives us confidence in the new results.

Our calculation shows the existence of three surface--induced bulk
states which have no dispersion from $\Gamma$ to $X$ and which do not change
in energy with different atom termination of the sample. These states
were first found experimentally
by Niles and H\"ochst\cite{niles} in CdTe(001). There
is also a distinct surface state for each the anion and the cation
termination of the surface. They do differ noticely in energy. A
general pattern develops for the location of the different bands in
II-VI wide band gap zincblende semiconductors within the range of their
valence band energy.

The rest of the paper is organized as follows. In Section II, to make
the paper self--contented,
we briefly
describe the main highlights of the method and present the formulae
that we have used. In Section III, we first
introduce the general characteristics of the valence band and the
location in energy of the surface and surface--induced bulk states. Next, we
discuss the results for each system studied. In a final section, we present our
conclusions.
We have included an appendix where we quote all the
tight--binding parameters of interest for this calculation.

\section{Method}
We make use of tight--binding Hamiltonians. Since the Green's function
matching method takes into account the perturbation caused by the surface
exactly, at least in principle, we can use the tight--binding parameters (TBP)
for the bulk.\cite{noguera,quintanar,rafa} This does not mean that we are
using the same TBP for the surface and the bulk.
Their difference is taken into account through the matching of the
Green's functions. We use the method in the form cast by
Garc\'\i a--Moliner and Velasco.\cite{g-m} They make use of the transfer
matrix approach first intoduced by Falicov and Yndurain.\cite{falicov}
This approach became very useful due to the quickly converging
algorithms of L\'opez--Sancho {\it et al.}\cite{sancho} Following the
suggestions of these authors, the algorithms for all transfer matrices needed
to deal with these systems can be found in a straightforward
way.\cite{trieste}
This method has been employed successfully for the
description of surfaces,\cite{daniel1,noguera,rafa}
interfaces,\cite{quintanar,rafa-prb}
and superlattices.\cite{arriaga,rafa-moliner}

\subsection{The formalism}
We have first, calculated the bulk (infinite medium) band structure of
the compounds by the tight--binding method (TB) in the Slater--Koster
language\cite{slater} using an orthogonal basis of five orbitals,
$sp^3s^*$. The $s^*$ state is introduced to properly locate in energy
the conduction band usually formed by $d$ states in the II--VI
zincblende (ZB) semiconductor compounds.\cite{austria,harrison} In our
calculation, we have included the effect of the spin--orbit (SO)
interaction.\cite{chadi}
The TBP that we have used in our calculation are listed in the
appendix. They reproduce the known bulk bands quite
well.\cite{daniel1,bertho,ang-cdse} We
assumed ideal truncation.

We obtain the Green's function from

\begin{equation}
\label{uno}
(\omega -H)G=I
\end{equation}
where $\omega$ is the energy eigenvalue, $H$ the tight--binding
Hamiltonian and $I$ is the unit matrix.
We adopt the customary decription in terms of principal layers. We
label them with positive numbers and zero for the surface principal layer.
Atomic layers are labelled with negative numbers and zero for the atomic
surface layer.
Let $\mid n\rangle$ be the principal wave function describing
the $n^{\rm th}$ principal layer. It is a LCAO wave function with
one $s$--like, three $p-$like, and one $s^*$--like atomic functions
per spin on each atom in the unit cell (there are two different atoms
in the cell, and two atomic layers per principal layer,
i. e., it is a 20--dimensional
vector). If we take matrix elements of eq. (\ref{uno}) in the Hilbert space
generated by the complete set of the wave functions $\mid n
\rangle$, we get
\begin{equation}
\label{dos}
\langle n \mid(\omega-H)\mid m \rangle = \delta_{mn}.
\end{equation}

Since there is only nearest--neighbors interactions between principal
layers, the identity opertator for the $n^{\rm th}$ principal layer is
\begin{equation}
\label{tres}
I=\mid n-1\rangle\langle n-1\mid + \mid n\rangle\langle n \mid +
\mid n+1\rangle\langle n+1\mid
\end{equation}
and therefore $H_{m,m+i}\equiv 0$ for $i\geq2$. By inserting eq.
(\ref{tres})
into eq. (\ref{dos}) we get
\begin{equation}
\label{cuatro}
(\omega -H_{nn})G_{nm}-H_{nn-1}G_{n-1m}-H_{nn+1}G_{n+1m}=
\delta_{mn}
\end{equation}
The matrix elements of the Hamiltonian, $H_{nm}$, that appear in this
formula are 2$\times$2 supermatrices (each principal layer contain two
atomic layers) each of whose elements is a 10$\times$10 matrix (since
we are using a 5--wave functions basis by spin per atom). For example,
for the surface
\begin{mathletters}
\begin{equation}
H_{00}=\left(
\begin{array}{lr}
	h_{00} & h_{0-1} \\
	h_{-10}& h_{-1-1} \\
\end{array}\right),
\end{equation}
\begin{equation}
H_{01}=\left(
\begin{array}{lr}
	h_{0-2} & h_{0-3} \\
	h_{-1-2}& h_{-1-3} \\
\end{array}\right).
\end{equation}
\end{mathletters}
Notice that rows are labeled with the index of the surface principal
layer zero (containing atomic layers 0 and --1) while the columns are
indexed with the first principal and second principal
layer (atomic layers 0 and --1 and atomic layers --2 and --3),
respectively.
We shall adopt the hypothesis of an ideal, non reconstructed surface.
Then, for the (001)--surface, for example, we have one atomic layer
of anions and one of cations per principal layer. Therefore
in this case $h_{00}\neq h_{-1-1}$, $h_{0-1}=h_{-10}^\dagger$. Thus,
to calculate $H_{00}$ and $H_{01}$ we need to know $h_{00},\ h_{-1-1},\
h_{0-1},$ and $h_{-1-2}$. These matrices are readily written in a
tight--binding language and can be calculate with the bulk parameters
as mentioned above. They depend on the wave vector {\bf k}.

Using eq. (\ref{cuatro}) for $n=m$, and $m=0$ for the surface, it is
straightforward to get the surface Green's function\cite{g-m}
\begin{equation}
\label{seis}
G_s^{-1}=(\omega I -H_{00})-H_{01}T
\end{equation}
and the principal layer projected bulk Green's function\cite{g-m}
\begin{equation}
\label{siete}
G_b^{-1}=G_s^{-1}-H_{10}\widetilde T
\end{equation}

It is customary to define the transfer matrices as
\begin{mathletters}
\begin{equation}
G_{k+1p}=TG_{kp},\qquad  G_{k+1p}=G_{kp}S, \qquad k\geq p\geq 0
\end{equation}
\begin{equation}
G_{ij+1}=\widetilde TG_{ij},\qquad G_{ij+1}=G_{ij}\widetilde S,
\qquad j\geq i\geq 0.
\end{equation}
\end{mathletters}
These matrices can be calculated by the quick algorithm of L\'opez--Sancho
{\it et al.}\cite{sancho}, recalculated later by
Baquero\cite{trieste} (see Refs. \cite{noguera,quintanar} for a
compilation of all the formulae).

{}From the knowledge of the Green's function, the surface states and
the surface--induced bulk states can be calculated from the poles
of the real part of the corresponding Green's function.
We have
applied previously this formalism to other
surfaces\cite{daniel1,noguera,rafa,angela}.
We will now present our results.

\section{Results}
The general characteristics of the zincblende II--VI wide band gap
semiconductor
valence band as we have obtained from our calculations is as follows.
The heavy--hole (hh) and light--hole (lh) bands follow each other closely in
energy. The corresponding wave functions are mainly\
($p_x ,\ p_y$) in character.
The hh band disperses from $\Gamma$ to $X$ about 2.0 eV and the lh one about
2.4 eV. The spin--orbit splitting is around 1 eV in the Te--compounds and
around 0.5 in the Se--ones. This band reaches
$X$ at about 5.0 eV. This is the band with the most  dispersion and is composed
essentially of $p_z$ states only. Finally, a very deep bulk band of
mostly $s-$character appears below --10 eV. It disperses about 2 eV
from $\Gamma-X$. We will call this band $b_{10}$.

{}From the (001)--bulk--projected Green's function we get also the
energy of the (001)--surface--induced bulk states. Three such states appear,
$B_h$, $B_l$ and $B_s$. These  surface--induced bulk bands show
no dispersion\cite{daniel1,prb2,daniel}. $B_s$ was first found
experimentally by Niles and H\"ochst\cite{niles} and confirmed later by
Gawlik {\it et al.}\cite{gawlik} for CdTe(001). $B_h$ mixes with the hh band at
$X$ and is located at the same energy (in $\Gamma$) than the surface state,
($S_c$, see below) of the cation terminated (001)--surface.
$B_l$ mixes with the lh band at $X$. The composition of both $B_h$ and $B_l$
is mainly of ($p_x ,\ p_y$), while the one of
$B_s$ which mixes with the spin--orbit band at $X$ is ($s ,\ p_z$).
The three states appear at the same position in energy irrespective of
the cation or
anion termination of the surface as one expects for surface--induced
{\it bulk} states which only depend on the surface through the boundary
condition (the wave function has to be zero at the surface).

The (001)--surface valence
band is rich\cite{daniel1,prb2} in other features. In particular,
three characteristic surface states do exist in this
range of energy. Two correspond to the anion
($S_{a1},\ S_{a2}$) and the other to the cation ($S_c$)
termination of the (001)--surface. In all the systems considered, the
anion terminated surface higher band ($S_{a1}$)
follows roughly the dispersion of the hh
bulk band but it
is at slightly higher energy. The cation terminated surface
band ($S_c$) starts roughly around 2--3 eV from the top of
the valence band in $\Gamma$ and has a varying ammount of dispersion.
The two states appear at very different energy values and are distinctive of
the termination of the surface for the four systems under consideration. One
could speculate on the use of these two states to characterize the termination
of the surface. A second (001)--surface anion state ($S_{a2}$) appears at
much lower energies, for all the compounds considered, near the lowest
bulk band $b_{10}$. This surface state has very little dispersion
and its wave function is mostly $s-$character.

There is only one set of experimental results available at this moment to
compare our results with, namely the already cited
CdTe(001).\cite{niles,gawlik}
We have studied this state and found that this
is a surface--induced bulk one. Such
bulk states appears as a consequence of the breaking of the symmetry
implied by the creation of the (001)--surface in the infinite medium. To
support
our argument, we showed\cite{daniel1}
specifically that a pole in the real part of the
(001)--bulk-projected
Green's function appears at --4.4 eV from $\Gamma$ to $X$, and
does not appear neither in the real part of the
(001)--surface--projected Green's function nor in the real part of
the bulk Green's function for any termination (anion or cation) of the
sample.
In other words, this is neither a solution of the Shrodinger's equation
for an infinite medium nor a solution for a semi--infinite one that is
localized
at the surface. It is a bulk state existing as a consequence of
the creation of the surface for which the crystal momentum is not a good
quantum number anymore due to the new boundary condition.
It has no $\epsilon(k)$--dispersion as a consequence. This is a
surface--induced
bulk state, as we already stated above.

In the rest of this paper, we want to describe in detail these
(001)--surface and (001)--surface--induced bulk states for the
four systems under consideration as well as the bulk bands obtained from our
calculation.
We will use our Hamiltonians to describe
interfaces, superlattices and quantum wells of these compounds in future work.

We summarize our results in Tables I--IV.  In Table I, we give the
characteristics of the bulk bands according to our obtained values. We give
the gap and the spin--orbit coupling for each element and the value for the
bands in $\Gamma$ and at $X$. In Table II, we give the energy of the three
surface--induced bulk states, then, in Table III, we give the energy of the
surface states
and, finally, in Table IV, the composition of all the bands
involved in our work.

\subsection{CdTe(001)}

Fig.1 shows the full valence band for this system.
The bulk states appear as the dotted curves in the figure.
Our tight-binding description reproduces correctly the band--gap value (1.6 eV)
and the spin--orbit splitting (0.9 eV). Our tight--binding parameters were
adjusted to give energy values at the $X$--high symmetry
point of the Brillouin zone that reproduce the experimental values
for CdTe very closely.\cite{daniel1,daniel}
The heavy hole
and light hole bands have a width of 1.8 eV and 2.2 for this
interval of the Brillouin zone, respectively. The wave functions are
constituted
mainly by an admixture of $p_x$ and $p_y$ states. See Tables I and IV.

The surface states
existing in this energy interval are identified as $S_{a1},\ S_{a2}$
and $S_c$. The
band $S_{a1}$ corresponds to the anion termination of the surface
(black triangles), its width $(\Gamma-X)$ is about 1 eV.
The wave functions are constituted mainly by an admixture
of $s$ and $p_z$ cation orbitals. The $S_c$ surface state (white triangles)
corresponds to the cation terminated (001)--surface. This band develops from
about 2.2 eV below the top of the valence band in $\Gamma$ to about --3.5 eV at
$X$. Its width is therefore about 1.3 eV and is mainly of $s$--character.
This band
mixtures with a surface-induced bulk band, ($B_l$), for about half of
the wave-vector interval $\Gamma-X$ (see below).
The $S_{a2}$ surface state shows a small dispersion and is located
at --8.5 eV. Its wave function is of $s-$character.
These surface states were described before.\cite{angela}

Three surface--induced bulk states exist in this range of energy according to
our findings. They are denoted by $B_h$, $B_l$ and $B_s$.

The states $B_h$ and $B_l$ are newly found surface--induced bulk states.
$B_h$ is located at --1.8 eV and has no dispersion as is characteristic
of these surface--induced bulk bands. $B_h$ seems to mix with the hh band at
$X$. $B_l$ appears at --2.2 eV and mixes with the lh band at $X$.
$B_l$ could
present branches with slighly different energy and be actually a group of
states differing slightly in energy. This states are mainly ($p_x,\ p_y$).
$B_l$ starts at
$\Gamma$ at the same energy as the surface state $S_c$.
Their energy difference is
very small for about half the interval $\Gamma-X$ when the surface state
begins to have an important dispersion while the bulk one has none.
We have obtained each
state from the corresponding Green's function.
The $B_s$
state has a $(s,\ p_z)$--composition.

In conclusion, CdTe(001) presents in the range of energy covered by its
valence band, as it is shown in Fig. 1, in adition to the known bulk bands
distinct surface states for each the anion ($S_{a1},\ S_{a2}$) and
the cation ($S_c$)
terminated (001)--surfaces and a total of three surface--induced bulk
states, $B_h$, $B_l$, and $B_s$. $B_s$ is the recently stablished
surface--induced
bulk state at --4.4 eV.\cite{niles,gawlik,daniel1}
None of them presents dispersion and they appear at
the same energy irrespective of the termination of the (001)--surface.

\subsection{CdSe(001)}
In general, CdSe is grown in an hexagonal structure. Nevertheless, recent
experiments\cite{ichino}
have shown that it is possible to grow this material in a cubic phase and
that it is stable. Our calculation has been done assuming the zincblende
structure. Our bands do not apply for the hexagonal structure.

We can see in Fig. 2 the general characteristic of the II--VI zincblende
wide band gap semiconductor valence band already described above.
The hh and lh bands
follow each other closely in energy and are both composed by
($p_x ,\ p_y$). The hh band
disperses from $\Gamma$ to $X$ 2.2 eV and the lh one 2.4 eV. The spin--orbit
splitting is 0.43 eV and the band ($p_z$) ends at --5.0 eV at $X$ and has
therefore a width of about 4.6 eV.
The $b_{10}$ band appears at --11.0 eV in $\Gamma$ and has little
dispersion towards $X$.

For the anion terminated (001)--surface, there is one
surface state ($S_{a1}$) of ($s,\ p_z$)--composition which develops right from
the top of the valence band in $\Gamma$ to $X$ with a quite less dispersion
than the hh band. At $X$ it takes the value 1.0 eV. The
cation terminated (001)--surface shows a surface state ($S_c$) of
$s$--character
at a lower enery in $\Gamma$ (--2.2 eV) that disperses very little up to
half the $\Gamma-X$ interval but then disperses strongly and ends at
--4.0 eV at $X$. Its width in the interval is 1.8 eV.
The $S_{a2}$ band appears again very closely to the $b_{10}$ bulk one
and shows very little dispersion as it is the general behavior for
this surface band.

The three (001)--surface--induced bulk states ($B_h$, $B_l$ and $B_s$) appear
at --2.2 eV ($p_x,\ p_y$), --2.45 eV ($p_x,\ p_y$) and
--5.0 eV ($s,\ p_z$) irrespective of the cation or anion termination of the
surface.
The first two mix with the hh and lh bands at $X$, respectively, and
the third one mixes with the spin--orbit band at $X$. None of them
presents dispersion.

As we can conclude from the previous description, CdSe(001) follows
very closely the
general picture described above for the number, composition and relative
location of the surface and surface--induced bulk states within the valence
band energy interval. We will now show that the last two compounds studied
follow the same pattern.

\subsection{ZnTe(001)}
The full valence band for ZnTe(001) is presented in Fig. 3. It is clear that
the general pattern is followed. $S_{a1}$ is the highest band
in energy. It starts at $\Gamma$ and disperses 1.0 eV
as it reaches $X$. It is of ($s,\ p_z$) character. The hh and lh bands
(both ($p_x,\ p_y$)) follow each other and disperse 2.0
and 2.3 eV, respectively.
$B_h$ and $B_l$ both
mix with the hh and lh band, respectively, at $X$ and have the same
composition.
The $S_c$ state disperses 2.0 eV and mixes with the $B_h$ band at $\Gamma$.
It is
$s$--like character following the general rule. Finally, the $B_s$ state
is located at
--5.3 eV from the top of the valence band and is of ($s,\ p_z$)--character.
This
state seems to be composed of two very nearly lying in energy states.
The surface--induced bulk states present no dispersion and are found at the
same energy irrespective of the surface termination.\cite{prb2} There is no
experimental evidence known to us for this findings.
The two surface states were described before.\cite{angela}

\subsection{ZnSe(001)}
We present our results for this compound in Fig. 4. In all respects ZnSe(001)
follow the general picture for the states in the valence band energy interval.
The hh, lh and spin--orbit bands have the usual composition and behavior.
Three different surface states appear, two for the anion--surface
($S_{a1},\ S_{a2}$) and one for the cation--surface ($S_c$),
and three surface--induced bulk states with the same usual
characteristics.
The
surface states have been studied previously.\cite{pollman}
We have summarize the details in Tables I--IV.

\section{Conclusions}
In conclusion, we have studied the (001)--surface and surface--induced
bulk states that
appear in the range of the valence band energy for the II--VI wide band gap
zincblende semiconductors. We found that a general pattern applies to the
states and relative energy of these states with respect to the bulk bands.
First, near the top of the valence band a surface state develops with
little dispersion from $\Gamma$ to $X$ for the anion--terminated surface.
A surface state also exists for the cation terminated surface but it
appears at a quite lower energy. A second anion--surface state appears
in the range of energies of the botton of the valence band.
Simultaneously three (001)--surface induced
bulk states appear which show no dispersion and appear at the same energy
irrespective of the surface termination. These states mix with the hh, lh and
spin--orbit bands at $X$, respectively.

It would be interesting to study other directions of the surface as well
as interfaces, superlattices and quantum wells in different high--symmetry
directions to characterize the different states that this break of symmetry
due to the creation of a border might induce.

\newpage
\appendix
\section*{}
We present in this appendix the tigh-binding parameters that we used
for the systems studied in this work.

Tight--binding parameters used in our calculation. We used
the notation of Bertho {\it et al.}\cite{bertho}

\begin{tabular}{lrlrl}
\tableline
\tableline
 Parameter&CdTe$^{\rm a}$&ZnTe$^{\rm b}$& ZnSe$^{\rm b}$& CdSe$^{\rm c}$\\
\tableline
\tableline
	$E_s^a$&       -8.19210&-9.19000 & -12.42728 & -10.16740\\
	$E_p^a$&        0.32790& 0.62682 & 1.78236 & 1.03400 \\
	$E_s^c$&       -0.95000& -1.42000 & 0.04728 & 1.07977   \\
	$E_p^c$&       6.93790 & 3.77952 & 5.52031 & 7.64650  \\
	$V_{ss}$&      -5.00000& -6.64227 & -6.50203 & -2.89240 \\
	$V_{xx}$&       2.13600& 1.94039 & 3.30861 & 3.01320  \\
	$V_{xy}$&       4.52817 & 4.07748 & 5.41204 & 5.73040    \\
	$V_{sp}^{ac}$& 3.31200 & 5.92472 & 1.13681 & 2.16040 \\
	$V_{sp}^{ca}$& 3.63824 & 4.67265 & 5.80232 & 5.65560 \\
	$V_{s^*s^*}^a$&10.44540& 6.22682 & 7.84986 & 6.02650  \\
	$V_{s^*s^*}^c$& 6.62960& 6.77952 & 8.52011 & 3.96150  \\
	$V_{s^*p}^{ac}$&2.52468& 2.96202 & 3.26633 & 2.11640 \\
	$V_{s^*p}^{ca}$&2.94540& 3.82679 & 1.86997 & 2.21680 \\
	$\lambda_a$&    0.32267& 0.36226 & 0.19373 & 0.14300  \\
	$\lambda_c$&    0.07567& 0.02717 & 0.01937 & 0.06700  \\
\tableline
\tableline
\end{tabular}

\noindent $^{\rm a}$ Ref. \cite{daniel1} \\
$^{\rm b}$ Ref. \cite{bertho} \\
$^{\rm c}$ Ref. \cite{ang-cdse}

\newpage

\begin{figure}
\caption{
The full electronic valence band structure for
CdTe(001) in the $\Gamma-X$ direction. The dot lines are the bulk
band structure from direct diagonalization of the tigh--binding Hamiltonian
using the tight--binding parameters that appear in the Appendix.
The two
surface states for the anion terminated (001)--surface are denoted by
full triangles, $S_{a1}$
and $S_{a2}$. The empty triangles are the surface state for
the cation terminated surface, $S_c$.
The (001)--surface--induced bulk
states, $B_h$, $B_l$ and $B_s$ are the full dots
(the dashed lines are keep up only for eye guide). Their occurence is
explained in the text. $B_s$ was first found experimentally by Niles
and H\"ochst.[3]
}
\label{figura1}
\end{figure}

\begin{figure}
\caption{
Electronic structure of the CdSe(001) valence band in the $\Gamma-
X$ direction. The conventions are the same as in Fig. 1 (See figure
caption). The (001)--surface--induced bulk
states, $B_h$, $B_l$ and $B_s$ mix at $X$ with the hh, lh and spin--orbit bands
respectively as a general rule. The pattern presented by the surface and
surface--induced bulk states as they appear in this figure is the general
one found also in the rest of the systems studied.
}
\label{fig2}
\end{figure}

\begin{figure}
\caption{
Electronic structure of the ZnTe(001) valence band in the $\Gamma-
X$ direction. See figure captions 1 and 2 for details.
}
\label{fig3}
\end{figure}

\begin{figure}
\caption{
Electronic structure of the ZnSe(001) valence band in the $\Gamma-
X$ direction. See figure captions 1 and 2 for details.
}
\label{fig4}
\end{figure}

\newpage

\begin{table}
\caption{
Characteristics of the band structure. $E_g$ is the gap
and $E_{so}$ is the spin--orbit splitting. The last three columns
are the values at the $X$--point of the Brillouin zone of the energy
of the heavy hole band ($E_{hh}$), the ligh hole band ($E_{lh}$),
and the spin orbit ($E_{so}$).
All the values are in eV.
}
\begin{tabular}{cccl||lcc}
\multicolumn{4}{c||}{} & \multicolumn{3}{c}{$X$--point}\\
\hline 
  & $E_g$ & $E_{so}$ & & $E_{hh}$ & $E_{lh}$ & $E_{so}$ \\
\tableline
\tableline
CdTe& 1.602 & --0.9 && --1.7 & --2.2 & -4.4 \\
CdSe& 1.78 &  --0.4 && --2.18& --2.36&--4.89 \\
ZnTe& 2.39 &  --0.91&& --1.93 & --2.40 & --5.50 \\
ZnSe& 2.82 &  --0.45&& --1.95 & --2.19 & --5.30 \\
\end{tabular}
\label{tabla1}
\end{table}

\vskip2.0truecm

\begin{table}
\caption{
The energy position of the (001)--surface--induced bulk states.
All the values are in eV.
}
\begin{tabular}{ccccc}
        & CdTe & CdSe & ZnTe & ZnSe \\
\tableline
$B_h$   & --1.8  & --2.2  & --2.0  & --2.0 \\
$B_l$   & --2.2  & --2.45 & --2.5  & --2.3 \\
$B_s$   & --4.4  & --5.0  & --5.5  & --5.3 \\
\end{tabular}
\label{tabla2}
\end{table}
\noindent

\vskip2.0truecm

\begin{table}
\caption{
The energy position of the (001)--surface states.
All the values are in eV.
}
\begin{tabular}{ccccl||lcc}
\multicolumn{5}{c||}{$\Gamma$--point}&\multicolumn{3}{c}{$X$--point}\\
\hline
        & $S_{a1}$ & $S_{a2}$ & $S_c$ & & $S_{a1}$ & $S_{a2}$ & $S_c$ \\
\tableline
CdTe   & 0.0  & --8.5 & --2.2  && --1.0  & --8.5 & --3.6 \\
CdSe   & 0.0  & --10.4 & --2.2  && --0.9 & --10.4 & --3.9 \\
ZnTe   & 0.0  & --11.6 & --2.0  && --0.9  & --11.7 & --4.4 \\
ZnSe   & 0.0  & --12.5 & --2.0  && --0.7  & --12.4 & --4.2 \\
\end{tabular}
\label{tabla3}
\end{table}

\vskip2.0truecm

\begin{table}
\caption{
The wave function decomposition for all the states that
appear in this study.
}
\begin{tabular}{cc}
State   & Composition \\
\tableline
\tableline
hh      & ($p_x,p_y$) \\
lh      & ($p_x,p_y$) \\
so      & ($p_z$) \\
\tableline
$S_{a1}$   & ($s,p_z$) \\
$S_{a2}$   & ($s$) \\
$S_c$   & ($s$)   \\
\tableline
$B_h$   & ($p_x,p_y$) \\
$B_l$   & ($p_x,p_y$)   \\
$B_s$   & ($s,p_z$)  \\
\end{tabular}
\label{tabla4}
\end{table}
\noindent

\end{document}